\documentclass[pra,twocolumn,superscriptaddress,showpacs]{revtex4-1}

\usepackage{graphicx}
\usepackage{bm}                 
\usepackage{amsmath}            
\usepackage{amssymb}
\usepackage{verbatim}           
\usepackage{amsthm}             
\theoremstyle{plain}            

\def\bra#1{{\langle#1|}}
\def\ket#1{{|#1\rangle}}

\def\expect#1{{\langle#1\rangle}}

\begin{document}

\title{Detection of discrete spacetime by matter interferometry}

\author{Todd A. \surname{Brun}}\email{tbrun@usc.edu}
\author{Leonard \surname{Mlodinow}}\email{lmlodinow@gmail.com}
\affiliation{Center for Quantum Information Science and Technology, University of Southern California, Los Angeles, California}

\date{\today}

\begin{abstract}
If the structure of spacetime is discrete, then Lorentz symmetry should only be an approximation, valid at long length scales. At finite lattice spacings there will be small corrections to the Dirac evolution that could in principle be experimentally detected. In particular, the lattice structure should be reflected in a modification of the free-particle dispersion relation. We show that these can produce a surprisingly large phase shift between the two arms of an asymmetrical interferometer. This method could be employed to test any model that predicts a direction-dependent dispersion relation. Here, we calculate the size of this phase shift for a particular model, the 3D quantum walk on the body-centered cubic lattice, which has been shown to give rise to the Dirac equation in the continuum limit. Though the details of this model will affect the size of the shift, its magnitude is set largely by dimensional analysis, so there is reason to believe that other models would yield similar results. We find that, with current technology, a modest-sized neutron interferometer could put strong bounds on the size of the lattice spacing. This discreteness could possibly be detected even for lattice spacings at the Planck scale by a suitably scaled-up experiment.
\end{abstract}


\maketitle

\section{Introduction}

\subsection{Discrete spacetime and Lorentz violation}

Lorentz symmetry, which underlies special and general relativity, has been tested to remarkable precision, and so far has passed every test.  This has not, however, prevented speculation that Lorentz might be violated at very high energies and very short length scales.  Over the years, a number of theories have been proposed that would lead to Lorentz violation, and a number of experiments have been suggested to test for such violations.

Dirac proposed that Lorentz invariance violation might play a role in physics in the 1950s \cite{Dirac51}, followed by several others in the years that followed (for example, \cite{Bjorken63,Pavlopoulos67,Redei67}). In the 1990s, influential papers by Coleman and Glashow raised the subject of systematic tests of Lorentz violation within the context of elementary particle physics \cite{Coleman97,Coleman99}. Various theories of quantum gravity have also suggested that Lorentz invariance may not be an exact symmetry \cite{Kostelecky89,Ellis99,Burgess02,Gambini99}. In those theories, the natural scale at which one would expect to observe that violation is at the Planck energy of approximately $10^{19}$ GeV, which would correspond to a distance scale of the Planck length, $M_{\rm Pl}=1.616\times10^{-35}$ m.  One type of Lorentz noninvariance that has been considered arises from discrete spacetime.  Some suggest that, at scales smaller than $M_{\rm Pl}$, the usual notions of space and distance may not even make sense \cite{Carr17}.

The Planck energy is not just far higher than current accelerator energies of approximately $10^3$GeV, but also far higher than the energy of the most energetic observed particles, the Ultra High Energy Cosmic Rays with energies as high as $10^{11}$ GeV $\approx 10^{-8} M_{\rm Pl}$. However, large violations of Lorentz invariance at the Planck scale can lead to a small degree of violation at much lower energies, presenting the possibility of detection at more realistic scales. For example, Lorentz violation could imply that neutrino oscillation will occur even if the neutrino mass is zero \cite{Kostelecky04}. Lorentz violation can also shift the threshold for elementary particle reactions, or lead to the occurrence of processes such as photon decay and the vacuum Cherenkov effect that would be forbidden in a Lorentz invariant theory.  As a result, the past few decades have seen a growing literature on tests of Lorentz invariance, and on the placement of bounds on a variety of proposed deviations (for recent reviews, see \cite{Mattingly05,Liberati13}).

One would expect the nature of proposed Lorentz violations to depend upon the specifics of the Lorentz-violating theory, but one common implication of such theories is an alteration of a particle's dispersion relation, which relates the particle's energy to its momentum and mass \cite{Mattingly05,Liberati13}. If spacetime has a discrete structure---such as an underlying lattice---these effects should be nonisotropic, giving different shifts in different directions.  A family of models that lead to such Lorentz violation are quantum walks and quantum cellular automata.

\subsection{Quantum walks}

Quantum walks---unitary analogues of classical random walks---were proposed both as possible constructions for quantum algorithms, and for simple models of quantum systems worthy of study for their own sakes \cite{AharonovY93,Ambainis01,AharonovD01,Kempe03,Brun03a,Kendon06}.  In a discrete-time quantum walk (the focus of the current work), a particle moves on a graph, the vertices representing possible particle positions and the edges the connections between them.  In addition to its position, such a particle has an internal space (often called the ``coin space'') which allows for nontrivial dynamics.

A body of work has shown that quantum walks on regular lattices in 1D, 2D and 3D can have as their continuum limit a relativistic quantum equation such as the Weyl and Dirac equations \cite{Bialynicki94,Meyer96,Strauch06,Bracken07,Chandrashekar10,Chandrashekar11,Chandrashekar13,DAriano14,Arrighi14,Farrelly14,DAriano15,Succi15,Bisio15,Arrighi15,Bisio16,Arrighi16,DAriano17,DAriano17b}.  In a previous work \cite{MlodinowBrun18}, we have shown that a particular construction of a 3D quantum walk on the body-centered cubic (BCC) lattice gives rise to the Dirac equation in that limit if it is invariant under parity transformations and discrete rotations of the coordinate axes.

According to that model, if spacetime were discrete the Dirac equation would be only approximately correct, and the degree to which nature deviates from the Dirac theory, and whether those deviations are observable, would be determined by the size of the lattice spacing $\Delta x$.  Experimental tests and astrophysical observations can therefore place upper limits on $\Delta x$.

The most glaring difference between the discrete and continuum theories is that Lorentz invariance is violated in the discrete theory \cite{Arrighi14b,Bisio17}.  In particular, the dispersion relation predicted by the quantum walk theory differs from that of the Dirac theory, in that the square of the energy, which is $m^2 c^4 + p^2 c^2$ for the Dirac theory, includes additional terms of order $k\Delta x$ and higher, which vanish in the continuum limit.  These higher-order terms can be seen as corrections to the continuum limit, which would act as perturbations to the usual Dirac evolution.

These terms have a directional dependence that could make them detectable in a suitably-designed (asymmetric) matter interferometer.  In this paper, we analyze the phase shift of such a matter interferometer, and calculate the magnitude of the shift for thermal neutrons.  The result, as we will see, is surprisingly large.  With interferometer sizes and accuracies typical of current neutron interferometers, one could put very stringent limits on the size of the lattice spacing.  Scaling up in accuracy by roughly three orders of magnitude would allow one to test for discreteness at the Planck scale.

We should emphasize that while the current calculations are specific to our particular 3D quantum walk model, such an experiment should also be able to see the effects of other theories with direction-dependent corrections to Lorentz invariance.  An experiment of this type would therefore test a whole family of Lorentz-violating models.

\subsection{This paper}

In section II of this paper we describe the 3D quantum walk on the BCC lattice, and derive its momentum-space representation.  In section III, we expand the evolution operator in the long wavelength limit, and equate that to the perturbation expansion of a Hamiltonian operator, which is the Dirac Hamiltonian to leading order, but has direction-dependent corrections at the next order.  This perturbation would produce spin- and direction-dependent energy shifts, which are calculated in section IV.  These shifts would in turn lead to phase shifts in an interferometer.

In section V.A we calculate the size of the relative phase shifts between the two arms of an asymmetric Mach-Zehnder interferometer.  This decomposes into the product of a quantity $(pmcL\Delta x/\hbar^2)$ that depends on the particle mass $m$, its momentum $p$, the linear size of the interferometer $L$ and the lattice spacing $\Delta x$, and a geometrical factor that depends on the layout of the interferometer and its orientation with respect to the underlying spatial lattice.  Though we calculate these quantities for our particular quantum walk model, the magnitude of the corrections is set largely by dimensional analysis, so one could reasonably expect that the size of the phase shifts predicted by other models would be comparable.  In section V.B we calculate the size of these phase shifts for a neutron interferometer with thermal neutrons, and estimate the limits that could be put on $\Delta x$ by current experimental abilities.  Finally, in section VI we discuss our results.

\section{3D Quantum Walk}

The unitary operator giving one step of the 3D quantum walk on the body-centered cubic lattice is
\begin{eqnarray}
U &=& \left( S_X P^+_X + S_X^\dagger P^-_X \right)
\left( S_Y P^+_Y + S_Y^\dagger P^-_Y \right) \nonumber\\
&& \times \left( S_Z P^+_Z + S_Z^\dagger P^-_Z \right) e^{-i \theta Q} ,
\label{eq:QWalk3DForm}
\end{eqnarray}
where $P^\pm_{X,Y,Z}$ are projectors onto the internal ``coin'' space, indicating whether the particle should step in the positive or negative direction along the $X$, $Y$, or $Z$ axis, and $S_{X,Y,Z}$ shifts the particle by one lattice position along the $X$, $Y$ or $Z$ axis.  This unitary operator $U$ has the form of three successive 1D quantum walks on the line, sharing the same internal space.  The operator $Q$ is the ``coin flip'' operator, which applies a rotation to the internal space.

We can most readily go to the continuum limit by transforming to the momentum representation:
\begin{widetext}
\begin{equation}
\ket{k_x,k_y,k_z} = \sum_{\ell,m,n=-\infty}^\infty e^{-i (k_x \ell\Delta x + k_y m\Delta x + k_z n\Delta x)} \ket{\ell\Delta x,m\Delta x,n\Delta x} ,\ \ \ \ -\pi < k_{x,y,z}\Delta x \le \pi .
\label{eq:3DMomentumStates}
\end{equation}
These are eigenstates of the shift operators $S_{x,y,z}$ with eigenvalues $e^{i k_{x,y,z} \Delta x}$.  In terms of this basis, the unitary evolution operator becomes
\begin{equation}
U = \int\int\int d^3\mathbf{k}\, e^{i k_x\Delta x \Delta P_X} e^{i k_y\Delta x \Delta P_Y} e^{i k_z\Delta x \Delta P_Z} e^{-i \theta Q} \otimes \ket{k_x,k_y,k_z}\bra{k_x,k_y,k_z} ,
\label{eq:ThreeDWalkMomentum}
\end{equation}
\end{widetext}
where $\Delta P_{X,Y,Z} \equiv P^+_{X,Y,Z} - P^-_{X,Y,Z}$, and $\Delta x$ is the lattice spacing.  As shown in earlier work \cite{MlodinowBrun18} these three $\Delta P_{X,Y,Z}$ operators and the coin flip operator $Q$ must all mutually anticommute.  Up to a unitary equivalence, these must therefore be the same as the operators in the Weyl representation of the Dirac equation:
\begin{eqnarray}
\Delta P_X &=& \gamma_0 \gamma_1 = - \sigma_Z \otimes \sigma_X , \nonumber\\
\Delta P_Y &=& \gamma_0 \gamma_2 = - \sigma_Z \otimes \sigma_Y , \nonumber\\
\Delta P_Z &=& \gamma_0 \gamma_3 = - \sigma_Z \otimes \sigma_Z , \nonumber\\
Q &=& \gamma_0 = - \sigma_X \otimes I .
\end{eqnarray}

We assume that one step of the quantum walk represents a time $\Delta t$, and define a limiting velocity $\Delta x/\Delta t \equiv c$.  The continuum limit is the limit where the wavelength is very long compared to the lattice spacing, which is $|k\Delta x| \ll 1$ for $k = \sqrt{k_x^2 + k_y^2 + k_z^2}$.  We also need the coin flip parameter $\theta$ to be small, so that in a single infinitesimal time step the coin flip operation is also infinitesimal.  Given a fixed finite $\Delta x$, the parameter $\theta$ is fixed by the particle mass:  $\theta \equiv mc\Delta x/\hbar$.

\section{Perturbation expansion}

As we approach the continuum limit, we can expand the unitary evolution operator in powers of $k\Delta x$.  The leading order (linear) term produces the Dirac equation.  But there are corrections to this evolution at quadratic and higher orders.  How can we consistently capture the effects of these corrections?

Our approach is to identify the time evolution operator with a Hamiltonian evolution, and then expand this Hamiltonian in a perturbation expansion.  That is, we equate
\begin{eqnarray}
U = \int\int\int d^3\mathbf{k} &\,& e^{-i(\Delta t/\hbar)(H_0(\mathbf{k}) + H_1(\mathbf{k}) + \cdots)} \nonumber\\
&& \otimes \ket{k_x,k_y,k_z}\bra{k_x,k_y,k_z} ,
\label{eq:HamiltonianExpansion}
\end{eqnarray}
where the $H_i$ operators act on the internal space, $H_0(\mathbf{k})$ is zeroth order in $k\Delta x$, $H_1(\mathbf{k})$ is first order, and so forth.  $U$ is given by Eq.~(\ref{eq:ThreeDWalkMomentum}).  We can expand both sides of Eq.~(\ref{eq:HamiltonianExpansion}) and equate them order by order.  At leading order we get the Dirac Hamiltonian:
\begin{equation}
H_0(\mathbf{k}) = -m c^2 Q + c\hbar\left( k_x \Delta P_X + k_y \Delta P_Y + k_z \Delta P_Z \right).
\label{eq:h0}
\end{equation}
Making the operator choices described above, and the usual identification of $\hbar \mathbf{k} = \mathbf{p}$, we get
\begin{equation}
H_0(\mathbf{k}) = -m c^2 \sigma_X\otimes I + c\sigma_Z \otimes \left( p_x \sigma_X + p_y \sigma_Y + p_z \sigma_Z \right).
\end{equation}
The eigenvalues of $H_0(\mathbf{k})$ are $\pm\sqrt{m^2 c^4 + c^2 p^2} \equiv \pm E_0$, where $p^2 = p_x^2 + p_y^2 + p_z^2$, and both eigenvalues are doubly degenerate.

At the next order in the expansion we get the equation
\begin{eqnarray}
(1/2)(\Delta t/\hbar)^2 H^2_0(\mathbf{k}) && \nonumber\\
+ (i\Delta t/\hbar) H_1(\mathbf{k}) &=& 
(1/2)(k^2 + \theta^2) I \nonumber\\
&& + \theta (k_x \Delta P_X + k_y \Delta P_y + k_z \Delta P_z) Q \nonumber\\
&& + k_x k_y \Delta P_X\Delta P_Y \nonumber\\
&& + k_x k_z \Delta P_X\Delta P_Z \nonumber\\
&& + k_y k_z \Delta P_Y\Delta P_Z .
\end{eqnarray}
Inserting our above expression for $H_0(\mathbf{k})$ and solving for $H_1(\mathbf{k})$, we get
\begin{eqnarray}
H_1(\mathbf{k}) = \frac{c\Delta x}{\hbar} && \biggl[ I\otimes (-p_y p_z \sigma_X + p_x p_z \sigma_Y - p_x p_y \sigma_Z) \nonumber\\ 
&& - mc \sigma_Y \otimes (p_x \sigma_X + p_y \sigma_Y + p_z \sigma_Z) \biggr] .
\end{eqnarray}
We could in principle continue the expansion to find higher orders of the perturbation series, but this first-order correction is already sufficient to produce observable consequences.  In particular, it produces energy shifts for free particles of equal momentum in different directions relative to the lattice.

\section{Energy shifts}

For a positive-energy eigenstate $\ket{v}$ of Hamiltonian $H_0(\mathbf{k})$ with energy $E_0 = \sqrt{m^2 c^4 + p^2 c^2}$, the next term in the perturbation expansion will produce a shift in energy
\begin{equation}
\Delta E_v(\mathbf{k}) = \bra{v} H_1(\mathbf{k}) \ket{v} .
\end{equation}
We can find the eigenvectors of $H_0(\mathbf{k})$ and calculate this energy shift exactly.  Let us rewrite this Hamiltonian as follows:
\begin{eqnarray}
H_0(\mathbf{k}) &=& -m c^2 \sigma_X\otimes I + c\sigma_Z \otimes \left( p_x \sigma_X + p_y \sigma_Y + p_z \sigma_Z \right) \nonumber\\
&=& -m c^2 \sigma_X\otimes I \nonumber\\
&& + pc\sigma_Z \otimes \left( \frac{p_x}{p} \sigma_X +\frac{p_y}{p} \sigma_Y + \frac{p_z}{p} \sigma_Z \right) \nonumber\\
&\equiv&  -m c^2 \sigma_X\otimes I + pc\sigma_Z \otimes \Phi_{\mathbf{\hat{p}}} ,
\end{eqnarray}
where
\begin{equation}
\Phi_{\mathbf{\hat{p}}} =  \left( \frac{p_x}{p} \sigma_X +\frac{p_y}{p} \sigma_Y + \frac{p_z}{p} \sigma_Z \right) .
\end{equation}
The operator $\Phi_{\mathbf{\hat{p}}}$ depends on the direction $\mathbf{\hat{p}}=\mathbf{p}/p$, and has eigenvalues $\pm1$ with corresponding eigenvectors $\ket{\phi_\pm}$.  Let us define its corresponding eigenvectors to be $\ket{\phi_\pm}$.  Then we can further rewrite $H_0(\mathbf{k})$ as:
\begin{eqnarray}
H_0(\mathbf{k}) &=& \left( -mc^2 \sigma_X + pc \sigma_Z \right) \otimes \ket{\phi_+}\bra{\phi_+} \nonumber\\
&& + \left( -mc^2 \sigma_X - pc \sigma_Z \right) \otimes \ket{\phi_-}\bra{\phi_-} \nonumber\\
&=& E_0\Biggl[ \left( - \frac{mc^2}{E_0} \sigma_X + \frac{pc}{E_0} \sigma_Z \right) \otimes \ket{\phi_+}\bra{\phi_+} \nonumber\\
&& + \left( - \frac{mc^2}{E_0} \sigma_X - \frac{pc}{E_0} \sigma_Z\right) \otimes \ket{\phi_-}\bra{\phi_-} \Biggr], \nonumber\\
&\equiv& E_0 \biggl[ \Psi_+ \otimes \ket{\phi_+}\bra{\phi_+} + \Psi_- \otimes \ket{\phi_-}\bra{\phi_-} \biggr] ,
\end{eqnarray}
where the operators $\Psi_\pm$ are
\begin{equation}
\Psi_\pm = \left( - \frac{mc^2}{E_0} \sigma_X \pm \frac{pc}{E_0} \sigma_Z \right)
\end{equation}
and both have eigenvalues $\pm1$, with corresponding eigenvectors $\ket{\psi_{\pm\pm}}$.  We can then identify two eigenvectors of $H_0(\mathbf{k})$ with positive energy $E_0$:
\begin{equation}
\ket{v_1} = \ket{\psi_{++}} \otimes \ket{\phi_+} ,\ \ \ \ \ 
\ket{v_2} = \ket{\psi_{-+}} \otimes \ket{\phi_-} ,
\end{equation}
and two eigenvectors with negative energy $-E_0$:
\begin{equation}
\ket{v_3} = \ket{\psi_{+-}} \otimes \ket{\phi_+} ,\ \ \ \ \ 
\ket{v_4} = \ket{\psi_{--}} \otimes \ket{\phi_-} .
\end{equation}
Restricting ourselves to the positive-energy eigenspace, we can then calculate the energy shift due to the first order perturbation Hamiltonian $H_1(\mathbf{k})$:
\begin{eqnarray}
\bra{v_1} H_1(\mathbf{k}) \ket{v_1} &=& -\left(\frac{c\Delta x}{\hbar}\right)\frac{p_x p_y p_z}{2p} , \nonumber\\
\bra{v_2} H_1(\mathbf{k}) \ket{v_2} &=& \left(\frac{c\Delta x}{\hbar}\right)\frac{p_x p_y p_z}{2p} , \nonumber\\
\bra{v_1} H_1(\mathbf{k}) \ket{v_2} &=& \left(\frac{c\Delta x}{\hbar}\right)
\sqrt{\left(\frac{m^2 c^2}{m^2 c^2 + p^2}\right) \left(\frac{p^2-p_z^2}{p^2}\right)} \nonumber\\
&& \times (p_x p_y - i p p_z)  \nonumber\\
&=& \bra{v_2} H_1(\mathbf{k}) \ket{v_1}^* .
\end{eqnarray}
We see that the energy shift produced by the perturbation depends on the {\it internal state} (spin) of the particle, and its {\it direction}.  These energy shifts should produce relative phase shifts between particles propagating in different directions.  This suggests in turn that a suitably arranged interferometer could, in principle, detect the effect of spacetime discreteness.

\section{Interferometer}

\subsection{Interferometer layouts}

To analyze the effect on an interferometer, we will treat the spatial degrees of freedom of the particle semiclassically, but the internal state of the particle quantum mechanically.  That is, we will assume that the particle is propagating in a wave packet that is broad enough in space to have a very narrow spread in momentum space about some central momentum $\mathbf{p}$, so that we can treat $\mathbf{p}$ as a definite value in solving for the evolution of the internal state.  So long as the uncertainty in momentum is very small compared to the momentum itself, the effect on the phase shift should be negligible.

Consider a wave packet with central momentum $\mathbf{p}$ propagating in a straight line over a distance $L$.  The time to traverse that distance is
\begin{equation}
t = \frac{L}{c} \sqrt{\frac{m^2 c^2 + p^2}{p^2}} .
\end{equation}
The state should accumulate a phase of
\[
\phi = - \frac{\expect{H_1(\mathbf{k})} t}{\hbar} .
\]
If the internal state of the particle is prepared in the eigenstate $(\ket{v_1}+\ket{v_2})/\sqrt2$ of the leading order Hamiltonian $H_0(\mathbf{k})$, then this phase becomes
\begin{eqnarray}
\phi &=& \left(\frac{L\Delta x}{\hbar^2}\right) \left(\frac{p_x p_y}{p}\right) \sqrt{\frac{m^2 c^2(p_x^2 + p_y^2)}{p^2}} \nonumber\\
&=& g(\mathbf{\hat{p}}) \left(\frac{pmcL\Delta x}{\hbar^2}\right)  ,
\label{eq:segmentShift}
\end{eqnarray}
where the geometric factor
\begin{equation}
g(\mathbf{\hat{p}}) \equiv \frac{p_x p_y \sqrt{p_x^2+p_y^2}}{p^3}
\end{equation}
depends only on the direction $\mathbf{\hat{p}}$ of the momentum vector, and the rest of the expression in Eq.~(\ref{eq:segmentShift}) depends only on its magnitude $p$.

\begin{figure}[htbp]
\begin{center}
\includegraphics[width=3in]{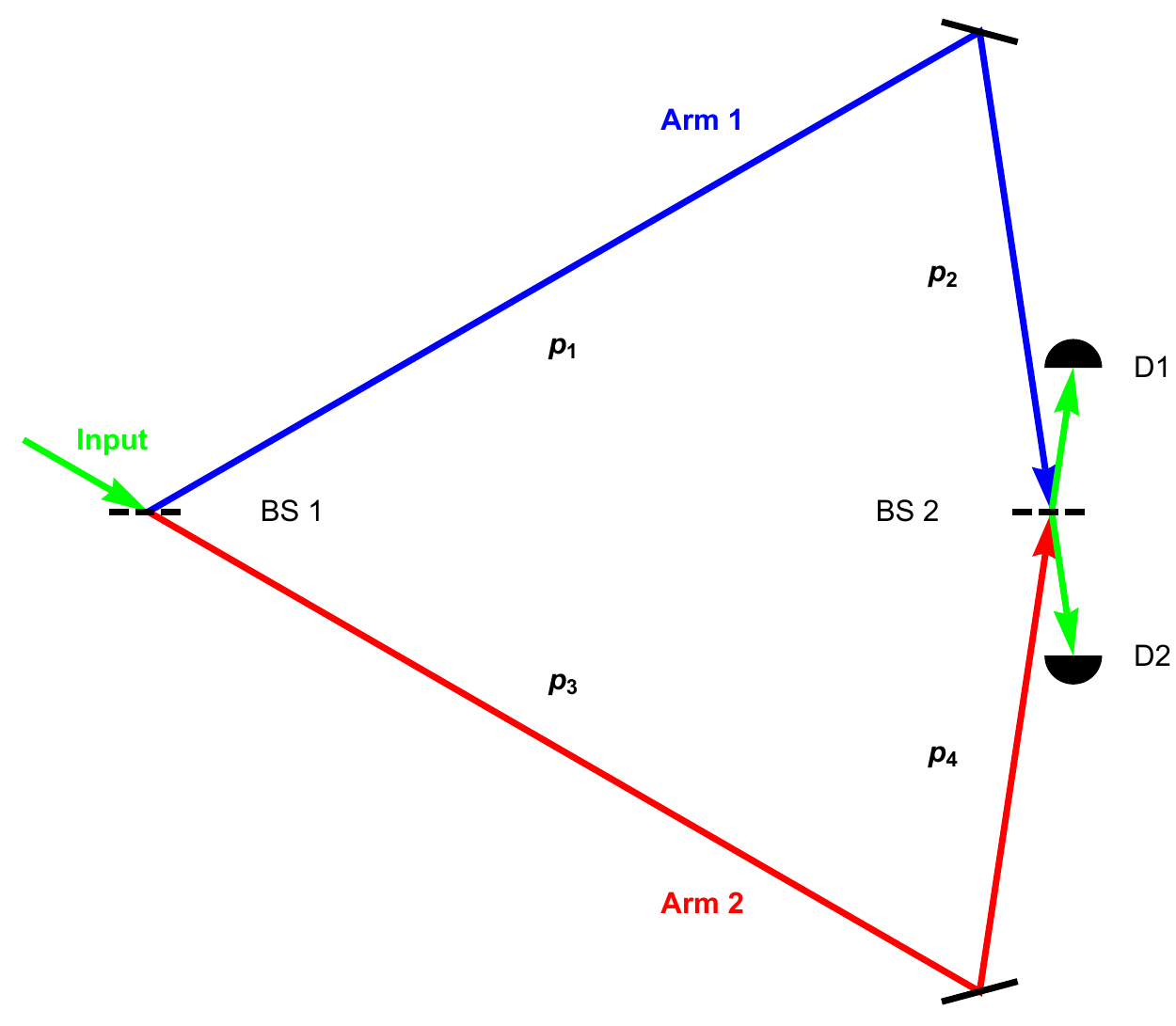}
\caption{An asymmetrical Mach-Zehnder interferometer that exhibits a relative phase shift between the two arms, which is dependent on its orientation relative to the spatial lattice and the spin state of the particle.  BS1 and BS2 are the beam splitters, and D1 and D2 are the detectors.}
\label{fig:interferometer1}
\end{center}
\end{figure}

We can now analyze the relative phase shift in an interferometer.  Suppose that the interferometer has two arms; each arm is a sequence of straight line segments.  The phase accumulated in one arm of the interferometer is the sum of the phases from each of the straight line segments, and the relative phase shift in the interferometer should then be the difference between the phases accumulated in each arm.  Because the phase depends on the direction of momentum, an asymmetric interferometer could accumulate different phases in each arm; and the relative phase will also generally depend on the orientation of the whole interferometer with respect to the underlying spatial lattice.

Consider an interferometer with an arrangement like that shown in Fig.~\ref{fig:interferometer1}.  The upper arm and lower arm are both of length $L$.  Each arm consists of a diagonal segment of length $2L/3$ and a near-vertical segment of length $L/3$.  Consider a 3D rotation
\begin{equation}
R(\theta_1,\theta_2,\theta_3) = R_X(\theta_1) R_Z(\theta_2) R_X(\theta_3) .
\end{equation}
If we rotate all the segments of the interferometer by this 3D rotation, we can calculate their contributions to the phase shift in each of the arms.  The relative phase shift between the upper and lower arms becomes
\begin{equation}
\Delta\phi \equiv \phi_1 - \phi_2 = g(\theta_1,\theta_2,\theta_3) \left(\frac{pmcL\Delta x}{\hbar^2}\right)   ,
\label{eq:relativePhase}
\end{equation}
where $g(\theta_1,\theta_2,\theta_3)$ is an overall geometric factor depending on the arrangement of the interferometer and its orientation with respect to the underlying spatial lattice.  For the interferometer depicted in Fig.~\ref{fig:interferometer1}, the factor $g$ is
\begin{eqnarray}
g(\theta_1,\theta_2,\theta_3) &=& (2/3) g(\mathbf{\hat{p}_1}) +  (1/3) g(\mathbf{\hat{p}_2}) \nonumber\\
&& -  (2/3) g(\mathbf{\hat{p}_3}) - (1/3) g(\mathbf{\hat{p}_4}) ,
\end{eqnarray}
and ranges from roughly $-0.57$ to $+0.57$.  In a random orientation it typically takes values with magnitude of order $10^{-1}$.

Note that we have no reason to believe that the interferometer in Fig.~\ref{fig:interferometer1} is optimal.  We have found one nonplanar interferometer arrangement, for example, which gives a geometrical factor that is even larger.  The interferometer depicted here has the largest geometrical factor $g$ of the small number of planar configurations we have analyzed, and large enough that it would not strongly suppress the signal in an experiment.

\subsection{Neutron interferometetry}

Neutrons have many properties that are useful for such an interferometry experiment.  They are uncharged fermions with high mass, so even thermal neutrons have relatively high momentum.  They can be prepared in spin-polarized input beams with a narrow range of momenta.  Physicists have a great deal of experience in designing and building high-precision neutron interferometers \cite{Zawisky98,Klepp14}.  Let's consider a neutron interferometer with a configuration like that depicted in Fig.~\ref{fig:interferometer1}.  The neutron has a mass $m = 1.675\times10^{-27}$ kg, and for thermal neutrons the momentum will be approximately $p = 3.7\times10^{-24}$ kg m/s.  A typical neutron interferometer has an arm length of order $L = 10$ cm.  Plugging these values into the formula in Eq.~(\ref{eq:relativePhase}) and assuming a geometric factor of order $g \sim 10^{-1}$, we get a relative phase shift of order $3\times10^{24} \Delta x$ radians between the arms (where $\Delta x$ is measured in meters).  The phase shift is directly proportional to the lattice spacing $\Delta x$.

This phase shift is surprisingly large.  A typical neutron interferometry experiment can resolve a phase shift on the order of $10^{-2}$ radians.  This would be large enough to detect a lattice spacing $\Delta x$ of about $3\times10^{-27}$ m.  To put that in perspective, the length scale probed at the LHC in CERN is about $10^{-18}$ m.  Moreover, the most accurate current experiments in neutron interferometry have a sensitivity down to microradians \cite{Cimmino89,Sarenac17}, which could put bounds on lattice spacings of order $10^{-31}$ m.

If the lattice spacing is of the order of the Planck length, $\Delta x \sim 1.6\times10^{-35}$ m, that would require a phase sensitivity on the order of nanoradians.  That is three or four orders of magnitude beyond the most accurate experiment to date.  However, the difference is not so large that such a measurement is beyond conception:  increasing the arm length to $\sim10$ m and collecting data for a long period could conceivably close that gap.

In practice, one would probably need to detect the change in relative phase between the two arms in different orientations of the interferometer, to distinguish the effects of Lorentz violation from small errors in the arm lengths of the interferometer or other such systematic uncertainties.  This raises some practical issues.  Because the interferometer is tied to an incoming beam line from a nuclear reactor, it probably cannot be rotated into an arbitrary orientation.  However, a small-scale interferometer (with arms of order 10 cm to 1 m) could be rotated azimuthally around the incoming beam line.

Of course, the interferometer is also on the surface of the earth, which is rotating.  Assuming that the earth's rotational axis has a fixed orientation relative to the underlying spatial lattice, one would expect the relative phase shift between the two arms to vary periodically with a period equal to the sidereal day (23 hours, 56 minutes and 4 seconds).  By collecting data over a long period, one could look for a periodic signal with that period.  One virtue is that systematic effects (due to daily temperature changes, for example) would more likely vary with the solar day.  Data could be collected for long periods for different azimuthal orientations of the interferometer and different spin states.  This could at the very least put a surprisingly stringent bound on the lattice spacing, or on other violations of Lorentz symmetry.

\section{Conclusions}

In this paper, we employed a model of discrete spacetime based on a 3D quantum walk on the BCC lattice, which has the Dirac equation as its long-wavelength limit.  Expanding the evolution operator in the small quantity $k\Delta x$, we found a perturbing Hamiltonian term that produces energy (and hence phase) shifts depending on both the direction (relative to the underlying spatial lattice) and the spin state of the particle.

Based on this Hamiltonian, we analyzed the design of an asymmetrical Mach-Zehnder interferometer that would exhibit a relative phase shift between its two arms that depends on its orientation.  Calculating the magnitude of this phase shift for spin-polarized thermal neutrons, we found that with current experimental accuracy, such a neutron interferometer could in principle put an upper bound on the lattice spacing $\Delta x$ of $10^{-27}$ m to as small as $10^{-31}$ m---only three or four orders of magnitude away from the Planck length, and vastly smaller than the length scales probed at the LHC.

While the numbers calculated in this paper pertain only to our particular quantum walk model, there is every reason to think that other models with discrete spacetime (or other direction-dependent violations of Lorentz invariance) would behave similarly.  The magnitude of the corrections is set largely by dimensional analysis; only the geometrical factors are likely to vary from model to model.  It would be interesting to compare such theories to each other, and estimate the range of effects they would produce in different interferometer configurations.  It is also possible that other kinds of matter interferometry might be even better than neutrons at detecting such Lorentz violating effects.  This also seems like a question worthy of exploration.

\begin{acknowledgments}

The authors acknowledge very useful information from Yuji Hasegawa, Dmitry Pushin, and Erhard Seiler, as well as interesting conversations with Chris Cantwell, Yi-Hsiang Chen, Shengshi Pang, and Chris Sutherland.  They are grateful for the hospitality of Caltech's Institute for Quantum Information and Matter (IQIM).

\end{acknowledgments}


\end{document}